\documentclass[fleqn,twoside]{article}
\usepackage{espcrc2}

\usepackage{graphicx}
\usepackage{epsfig}
\usepackage[figuresright]{rotating}

\newcommand{\AmS}{{\protect\the\textfont2
  A\kern-.1667em\lower.5ex\hbox{M}\kern-.125emS}}

\newcommand{\red}[1]{{\textsl{$#1$}}}

\def\aitalc{{\sc \textit{a}{\r{\i}}\raisebox{-0.14em}{T}alc}}
\def\hahn{\textrm{FA/FC/LT}}

\title{
\vspace*{-1.cm}
{\normalsize \tt SFB/CPP-04-03
\\ March 2004} 
\\
\vspace*{1.cm}
Automated use of DIANA for two-fermion production at colliders%
\thanks{Work supported in part by European
          Community's Human Potential Programme under contract
          HPRN-CT-2000-00149, by 
          Sonderforschungsbereich/Transregio 9-03 of DFG
          `Computergest\"utzte Theoretische Teilchenphysik', and
          by the Polish State Committee for Scientific Research (KBN)
          for the research project in 2004-2005.}}

\author{J. Gluza%
\address[DESYZ]{DESY, Platanenallee 6, 15738
          Zeuthen, Germany}%
\address[Katowice]{Institute of Physics, Univ. of
    Silesia, Universytecka 4, 40007 Katowice, Poland},
        A. Lorca\addressmark[DESYZ]
and
        T. Riemann\addressmark[DESYZ]\thanks{Corresponding author.
        DESY, Platanenallee 6, 15738
          Zeuthen, Germany,
         {\it E-mail address:} Tord.Riemann@desy.de (T. Riemann). }
}
       
\begin{document}

\begin{abstract}
We describe packages for the calculation of radiative corrections to
two-fermion production at colliders.
The packages use {\tt DIANA}, and also {\tt QGRAF}, {\tt FORM}, {\tt
  Fortran}, and further 
unix/linux tools. 
The one-loop calculations in the Standard Model are highly automatized
with the package \aitalc{}.
Further, the automatic determination of all the matrix elements for two-loop
corrections to massive Bhabha scattering in QED and the classification of their
topologies and prototypes is done with {\tt  DIANA}.
A generalization to the Standard Model is straightforward.

\vspace{1pc}
\end{abstract}

\maketitle

\section{INTRODUCTION}
For the future $e^+e^-$ Linear Collider (LC) we need quite precise
predictions of cross-sections for a variety of reactions, among them 
\begin{eqnarray}
\label{eq1}
 e^+e^- \to {\bar f} f(\gamma), ~~~~f=e,\mu,\tau,u,d,c,s,t,b .
\end{eqnarray}
The predictions have to be calculated with account of quantum
corrections, typically with one- or two-loop accuracy.
This has to be done in some model of field theory, notably the Standard
Model or the Minimal Supersymmetric Standard Model.
First complete one-loop calculations in the Standard Model for
two-fermion production \cite{Passarino:1979jh} and for Bhabha
scattering \cite{Consoli:1979xw} date back to 1979.
After 25 years of perturbative calculations, the quest for a more
or less complete automatization of this kind of calculation is natural.
Nevertheless, only few packages for such a task are publicly available.
We would like to mention {\tt CompHEP} 
\cite{Boos:1989dx,Pukhov:1999gg,Boos:2003acat,Kryukov:2003acata1,Kryukov:2003acata2},
{\tt FeynCalc/FeynArts/FormCalc/LoopTools/FF}
\cite{Kublbeck:1990xc,Hahn:2000kx,Hahn:1998yk,vanOldenborgh:1991yc2},
the {\tt grace} project \cite{Belanger:2003sd,Fujimoto:2003acat}
and {\tt SANC} \cite{Bardin:2003zd}.
These packages have a variety of options to be used, but they do not go
substantially beyond the one-loop level.

We are interested in a package for one- and two-loop calculations and
we think that a prospective approach might be based on {\tt DIANA}
\cite{Tentyukov:1999is,Tentyukov:2002ig,Fleischer:2003acat},
an interface to {\tt QGRAF} \cite{Nogueira:1993ex},
to be combined also with {\tt FORM}
\cite{Vermaseren:1991??,Vermaseren:2000nd} and 
{\tt Fortran}, {\tt c++} and additional unix/linux tools.  

Recently, we performed complete, high-precision one-loop
calculations for (\ref{eq1}) ($f\neq e$) in the Standard Model
\cite{Fleischer:2003kk,Hahn:2003ab} 
and demonstrated numerical agreements with the results of
other groups, finally with up to eleven digits \cite{Lorca:2003?1}.
In a next step, the program package was
completely rewritten in order to allow also the unexperienced user to
create their own numerical code.
The result is the package \aitalc.
It covers also Bhabha scattering and will be described in more detail
in the next section.
In an earlier comparison for Bhabha scattering 
\cite{Bardin:1991xe} a per mill agreement was established (see also 
\cite{Bardin:1997xq2,Kobel:2000aw}). 
For realistic applications, one has to include also higher order
corrections and to combine the so-called `weak library' with
a Monte Carlo code for the treatment of real bremsstrahlung; this is
not discussed here.
A dedicated comparison of this kind proved an accuracy of the order of
$10^{-4}$ \cite{Beenakker:1998fi}.
\section{DIANA and two-fermion production}
It is not so long ago that {\tt DIANA} \cite{Tentyukov:2004??} was in a kind
of experimental state.
There exist several variants of the package with identical version
numbers but with quite some different properties.
For this reason, we decided to install a chain of versions with
well-defined version properties at our location, presently the last
one being v.{}2.35
\cite{zeuthen-diana-235:2004}.
From the user's point of view, the package is characterized by a small
number of input files:
\begin{itemize}
\item {\tt process.cnf} -- the file defines how to run {\tt DIANA};
  for examples see \cite{zeuthen-diana-235:2004}; the incoming and
  outgoing particles plus loop order should be edited;
\item {\tt Model.mode}l -- we have prepared four model files:
      {\tt QED.model}, {\tt
        Standard\linebreak[2]Model0.\linebreak[2]model} (a basic file
      with leptons

      only), {\tt StandardModel.model} 
(complete with all masses and mixings), {\tt StandardModelC.model} (with
counter terms for neutral current $2f\to{2f}$, following
\cite{Denner:1993kt}, a typo corrected with \cite{Bohm:1986rj});  
\item 
{\tt FeynmanRules.frm} --  we have prepared three sets of Feynman rules: 
{\tt FeynmanRules0.frm} (a simple version),
{\tt FeynmanRules1.frm} (with account of actual external momenta distribution),
{\tt FeynmanRules2.frm} (with account of internal momenta).
\end{itemize}
These files, together with a proper application of {\tt QGRAF} and of
additional  {\tt DIANA} 
options, give a huge flexibility to prepare a sample of prepared
matrix elements of a given loop order in a given model for a
subsequent {\tt FORM} or {\tt MAPLE} calculation.
As an application in the Standard Model, the flavor non-diagonal
reactions  
\begin{eqnarray}
\label{eq2}
 e^+e^- \to {\bar f_1} f_2, ~~~~f_{1,2} = \mu,\tau,u,d,c,s,t,b 
\end{eqnarray}
may also be treated.
In this respect it would be extremely useful to have additional model
files,  e.g. for the minimal supersymmetric Standard Model.
Some time ago, a project in XML was proposed, being intended on
creating a standard for model files to be used in different CAS  
 \cite{Demichev:2002he}.
\subsection{Full automatization of one-loop EWRC: \large
  \protect \aitalc} 
\begin{figure}[ht]
\includegraphics[width=213pt]{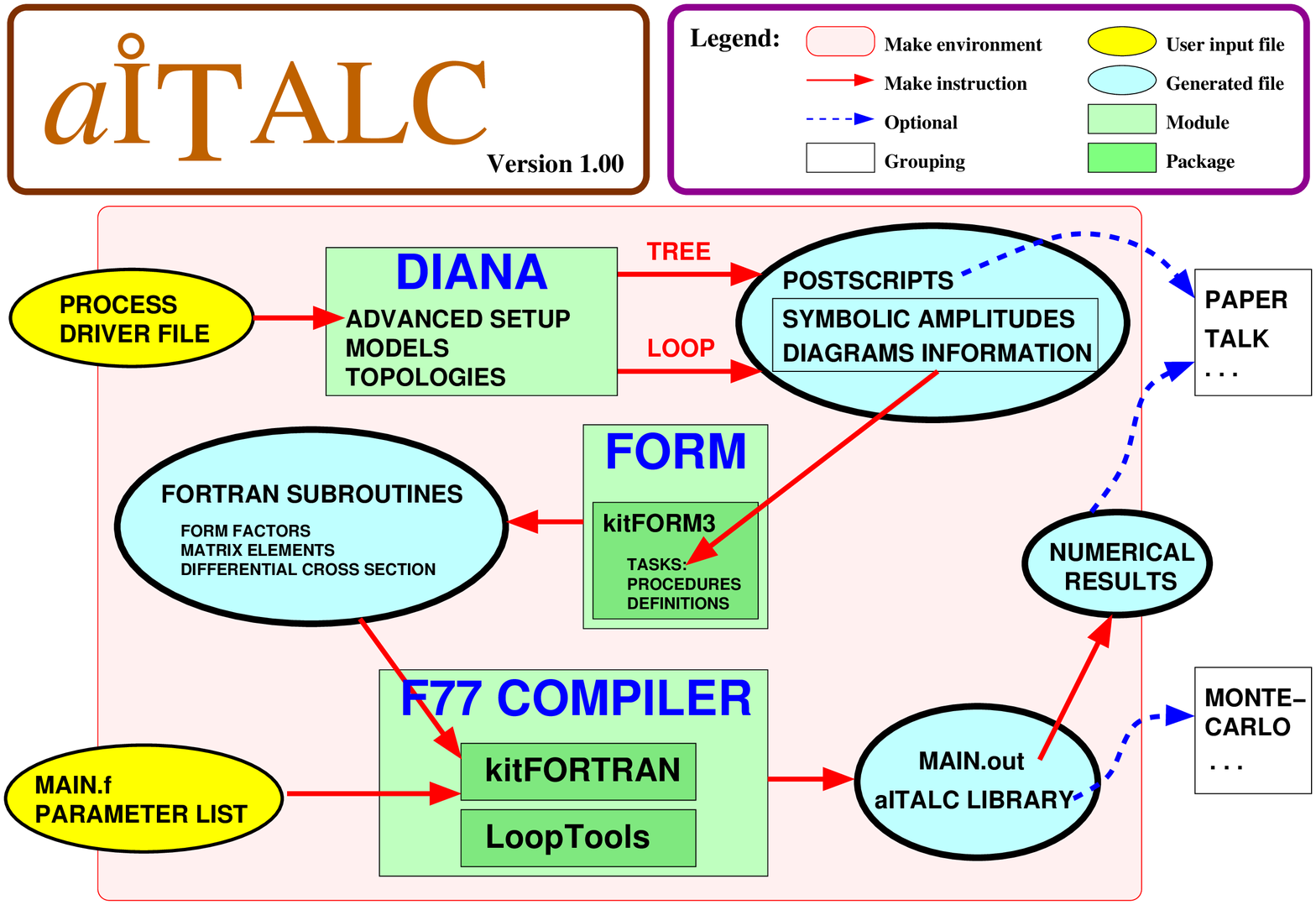}
\vspace{-5pt}
\caption{Flow chart of the \protect \aitalc{} package.}
\label{flowchart}
\end{figure}

A completely automated example, based on {\tt DIANA} and other
packages (see Fig. \ref{flowchart}), 
is the project {\aitalc} \cite{zeuthen-aitalc:2003},
that provides also numerical output.

It is designed to organize the
calculation of tree-level and one-loop corrections to $2 \rightarrow 2$ fermion
processes, including Bhabha scattering (and
in the near future also Moller scattering).
This tool allows presently the calculation of unpolarized
differential and integrated cross-sections.
Differing from {\tt topfit} \cite{Fleischer:2003aa}, {\aitalc} treats
all particle masses (including the electron mass) and mixings exactly,
but does not cover hard photon emission.

In a simple driver file {\tt process.ini} (simple form of {\tt
  process.cnf}), the user specifies the ingoing and outgoing 
fermions.
The determination of matrix elements, form factors and
other kinematical routines proceeds then automatically.
Later on, two Fortran files give access to
the specific values of the parameters in our model and the numerical
output of the code (i.e. a number of data points in the angular
distribution, integrated cross section, running flags, etc).
\\
Advanced features are not extremely user friendly, but still under
development. 
Fermion masses are included by default, but may be neglected, and
soft photonic corrections may be included (or not), as well as one may
perform a full one-loop tensor integral reduction to the master integrals  
$A_0$, $B_0$, $C_0$ and $D_0$ in the Passarino-Veltman scheme
\cite{Passarino:1979jh}. 
All these possibilities will be described in more detail in a tutorial
to be published soon.
\\
Some more details may be found also in
\cite{Lorca:2003?1}.
\section{NUMERICAL ONE-LOOP RESULTS FOR BHABHA SCATTERING}
The package \aitalc{} was used for a precise calculation of the one-loop
electroweak corrections to fermion pair production and Bhabha
scattering at LC energies.
We compared the result of this with numerics from
\mbox{FeynArts}/\mbox{FormCalc}/\mbox{LoopTools}
\cite{Lorca:2003?1}.

The Table \ref{numbersoneloop} shows an agreement of the two calculations of 14 digits for
the $\mathcal{O}(\alpha)$ corrections (i.e. Born+QED+weak+soft) for
the case with an exact treatment of $m_e \neq 0$, while the simplified
comparison with $m_e \to 0$ does not provide more than 10 digits agreement.
For practical purposes, this make no difference, of course.

\begin{table}[t]
\caption{Cross-sections for $e^- e^+ \to e^- e^+ ~(\gamma)$ at
$\sqrt{s}=500~\mathrm{GeV}$ with  a photon energy cut 
$E^{\mathrm{max}}_{\gamma_{\mathrm{soft}}}={\sqrt{s}}/{10}$.} 
\label{numbersoneloop}
$
\begin{array}{rll}
\hline
\cos{\theta} \phantom{\Big|}
& {\left[ \frac{\mathrm{d} \sigma}{\mathrm{d} \cos{\theta}} \right]}_{\mathcal{O}(\alpha^3)} \quad \mathrm{(pb)}
& \mathrm{Group} \\
\hline
  -0.9 
& 0.19344~50785~2686\red{3~6}\phantom{\cdot 10^0}& \textrm{\aitalc{}}\\
  -0.9 
& 0.19344~50785~2686\red{2~2}\phantom{\cdot 10^0}& \hahn\\
\vspace{5pt}
  -0.9 
& 0.19344~50785~\red{62637~9}\phantom{\cdot 10^0}&\red{m_e=0} \\

   0.0 
& 0.54667~71794~6942\red{3~1}\phantom{\cdot 10^0}& \textrm{\aitalc{}}\\
   0.0 
& 0.54667~71794~6942\red{1~8}\phantom{\cdot 10^0}& \hahn\\
\vspace{5pt}
   0.0 
& 0.54667~71794~\red{99961~4}\phantom{\cdot 10^0}&\red{m_e=0} \\


   0.9 
& 0.17292~83490~6650\red{7~2}\cdot 10^3& \textrm{\aitalc{}}\\
   0.9
& 0.17292~83490~6650\red{8~0}\cdot   10^3& \hahn \\
   0.9 
& 0.17292~83490~6\red{1347~4}\cdot 10^3 &\red{m_e=0} \\
\hline
\end{array}
$
\vspace*{-0.3cm}
\end{table}

\section{MASSIVE TWO-LOOP BHABHA SCATTERING}
\begin{figure}[t]
\epsfig{file=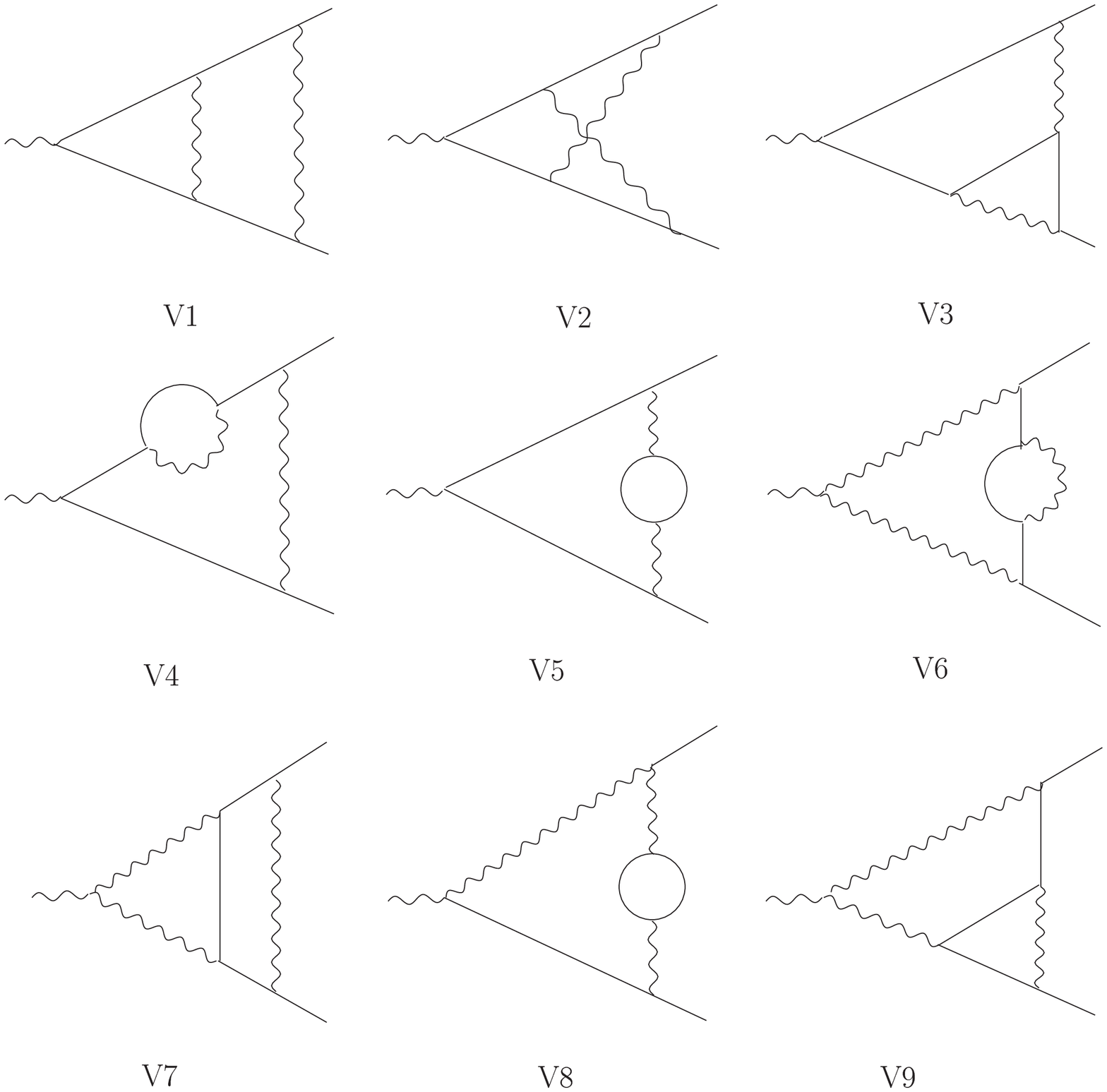, width=7.5cm}
\epsfig{file=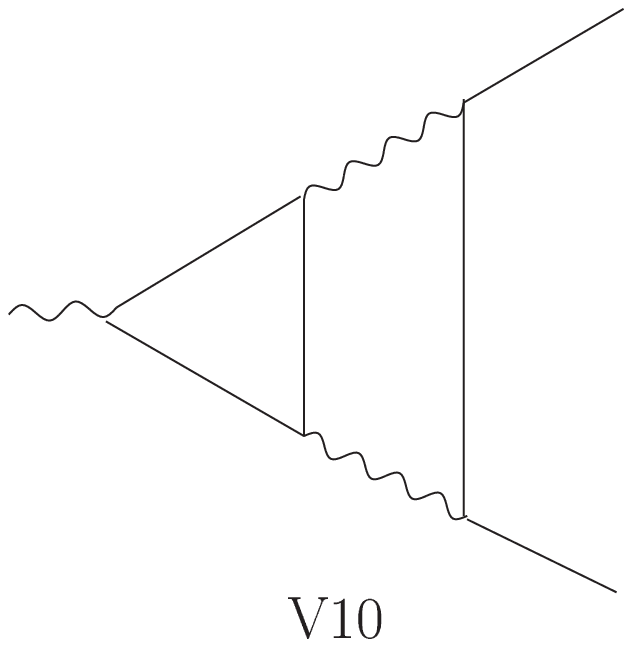, width=2.2cm}
\vspace*{-0.6cm}
\caption{Vertices for the two-loop Bhabha process with six internal lines.}
\label{acat-riemann-fig-1}
\vspace*{-0.4cm}
\end{figure}

In a related activity, we use {\tt DIANA} for a preparation of two-loop
matrix elements in massive QED.
We apply  it for the solution of two different problems:
\begin{itemize}
\item
The
calculation of 
interferences of the two-loop amplitudes with the Born amplitudes
(i.e. of the integrands for the determination of the scalar
integrals);
\item
The determination of all prototypes.
\end{itemize}
Prototypes are topologies of diagrams, where additionally the
different masses on internal lines are taken into account.
An example is shown in Figure \ref{acat-riemann-fig-1}, where all 
prototypes with three external lines and six internal lines for the Bhabha
process are given.  
Diagrams V1-V5 are genuine two loop QED vertices for the Bhabha process.
Vertices V6-V10 come from extractions of one single line from  
two-loop
QED box prototypes (altogether there are six genuine box prototypes).   
Strictly speaking,
the prototypes V4-V6 and V8 should already be shrinked to five internal
lines and  
classified into this  class of prototypes because they have
two internal lines with identical momentum. 
There are many more prototypes
with five internal lines which come from the extraction of one (or
two) lines of  
original vertex (or box) prototypes; we do not present them here. 
For these cases and for 
cases where even more lines are extracted, an automatization of the procedure 
for the identification of  prototypes is very useful.
Technically it has been  made by merging information from  
{\tt DIANA} and {\tt QGRAF} in a subsequent {\tt FORM} program.
 
Let us mention finally that  out of all the prototypes shown in Figure 
\ref{acat-riemann-fig-1}  only V2 and V10 correspond to master
integrals; all the others may be reduced to other master integrals. 
For a complete set of master integrals see \cite{Czakon:2004nn2}.


\section*{Acknowledgements}
We would like to thank M. Czakon, J. Fleischer and M. Tentyukov for close
collaboration in research related to this project and T. Hahn for
support of the numerical comparisons. 

{

\providecommand{\href}[2]{#2}

}

\end{document}